\documentclass[pre,twocolumn,showpacs,showkeys,superscriptaddress,floatfix,
amsmath,amssymb,a4paper]{revtex4}

\usepackage{graphicx}% Include figure files
\usepackage{dcolumn}% Align table columns on decimal point
\usepackage{bm}% bold math
\DeclareMathOperator{\sech}{sech}

\newcommand{\eq}[1]{Eq.~(\ref{#1})}
\newcommand{\fig}[1]{Fig.~\ref{#1}}         
\newcommand{\iotabar}{{\mbox{$\iota\!\!$-}}}
\newcommand{\iotadot}{\dot{\iotabar}}
\newcommand{\iotaddot}{\ddot{\iotabar}}
\newcommand{\iotadddot}{\dddot{\iotabar}}
\newcommand{\Ds}{D_{\mathrm{S}}}
\newcommand{\Dsdot}{\dot{D}_{\mathrm{S}}}
\newcommand{\Dsddot}{\ddot{D}_{\mathrm{S}}}
\newcommand{\dotv}{\mbox{\boldmath\(\cdot\)}}

\newcommand{\const}{{\mathrm{const}}}
\newcommand{\Bvec}{{\mathbf{B}}}
\newcommand{\Ften}{{\mathbf{\mathsf{F}}}}
\newcommand{\kvec}{{\mathbf{k}}}
\newcommand{\rvec}{{\mathbf{r}}}
\newcommand{\xivec}{{\bm{\xi}}}

\newcommand{\mmax}{m_{\mathrm{max}}}

\newcommand{\rmu}{r_{\mu}}

\newcommand{\half}{{\textstyle{\mathrm{\frac{1}{2}}}}}
\newcommand{\quarter}{{\textstyle{\mathrm{\frac{1}{4}}}}}

\begin{document}

\preprint{Submitted to Phys. Rev. E May 1, 2004}

\title{Statistical characterization of the interchange-instability spectrum of a separable
ideal-magnetohydrodynamic model system}

\author{R.L. Dewar}
\email{robert.dewar@anu.edu.au}
\affiliation{%
 Department of Theoretical 
 Physics and Plasma Research Laboratory, Research School of Physical Sciences and Engineering,
 The Australian National University, ACT 0200, Australia}
 \affiliation{%
 Graduate School of Frontier Sciences,
 University of Tokyo,
 5-1-5 Kashiwanoha, Kashiwa-shi, Chiba, Japan 277-8651
}
\author{T. Tatsuno}%
\affiliation{%
Institute for Research in Electronics and Applied Physics,
University of Maryland,
College Park, MD 20742-3511, USA
}%
\affiliation{%
Graduate School of Frontier Sciences,
University of Tokyo,
5-1-5 Kashiwanoha, Kashiwa-shi, Chiba, Japan 277-8651
}%
\author{Z. Yoshida}%
\affiliation{%
Graduate School of Frontier Sciences,
University of Tokyo,
5-1-5 Kashiwanoha, Kashiwa-shi, Chiba, Japan 277-8651
}%
\author{C. N\"uhrenberg}
\affiliation{%
Max-Planck-Institut f\"ur Plasma Physik,
Teilinstitut Greifswald, D-17491 Germany}
\author{B.F. McMillan}
\affiliation{%
 Department of Theoretical 
 Physics and Plasma Research Laboratory,
 Research School of Physical Sciences and Engineering,
 The Australian National University, ACT 0200, Australia}

\date{May 1, 2004}% It is always \today, today,

\begin{abstract}
		A Suydam-unstable circular cylinder of plasma with periodic
		boundary conditions in the axial direction is studied within the
		approximation of linearized ideal magnetohydrodynamics (MHD).  The
		normal mode equations are completely separable, so both the
		toroidal Fourier harmonic index $n$ and the poloidal index $m$ are
		good quantum numbers.  The full spectrum of eigenvalues in the
		range $1 \leq m \leq \mmax$ is analyzed quantitatively, using
		asymptotics for large $m$, numerics for all $m$, and graphics for
		qualitative understanding.  The density of eigenvalues scales like
		$\mmax^2$ as $\mmax \rightarrow \infty$.  Because finite-$m$
		corrections scale as $1/\mmax^2$, their inclusion is essential in
		order to obtain the correct statistics for the distribution of
		eigenvalues.  Near the largest growth rate only a single radial
		eigenmode contributes to the spectrum, so the eigenvalues there
		depend only on $m$ and $n$ as in a two-dimensional system.
		However, unlike the generic separable two-dimensional system, the
		statistics of the ideal-MHD spectrum departs somewhat from the
		Poisson distribution, even for arbitrarily large $\mmax$.  This
		departure from Poissonian statistics may be understood
		qualitatively from the nature of the distribution of rational numbers
		in the rotational transform profile.
\end{abstract}

% http://www.aip.org/pacs/pacs03/pacs0350.html
%Use showkeys class option if keyword
\maketitle

\section{Introduction\label{sec:Intro}}

The general aim of this paper is to compare and contrast the spectrum
of eigenvalues in typical integrable wave systems (e.g. waves in a
rectangular cavity \cite{casati-chirikov-guarneri85}) with the spectrum
of instabilities in a cylindrical plasma within the ideal
magnetohydrodynamics (MHD) approximation.  This is a first step in
understanding the spectral problem in the complex three-dimensional
geometry of the class of magnetic confinement fusion experiments known
as stellarators \cite{wakatani98}.

In ideal MHD the spectrum of the frequencies,
$\omega$, of normal modes of displacements about a toroidal
equilibrium is difficult to characterize mathematically because the
linearized force operator is not compact \cite{lifschitz89}.  In
addition to a point (discrete) spectrum of unstable modes ($\omega^2 <
0$) there are the Alfv\'en and slow-magnetosonic continuous spectra on
the stable side of the origin ($\omega^2 > 0$) and the possibility of
dense sets of accumulation points on the unstable side.  In
mathematical spectral theory the stable continua and unstable
accumulation ``continua'' \cite{spies-tataronis03} are characterized
\cite{hameiri85} as belonging to the \emph{essential
spectrum}.\ (For a self-adjoint operator $L$, the essential
spectrum is the set of $\lambda$-values for which the range of
$L-\lambda$ is not a closed set and/or the dimensionality of the null
space of $L-\lambda$ is infinite.)

There is experimental evidence that ideal MHD is relevant in
interpreting experimental results \cite{troyon_etal84,ferron_etal00},
but perhaps the greatest virtue of ideal MHD in fusion plasma physics
is its mathematical tractability as a first-cut model for assessing the
stability of proposed fusion-relevant experiments with complicated
geometries in the pre-design phase.

For this purpose a substantial investment in effort has
been expended on developing numerical matrix eigenvalue programs, such
as the three-dimensional TERPSICHORE \cite{anderson_etal90} and CAS3D
\cite{schwab93} codes.  These solve the MHD wave equations for
perturbations about static equilibria, so that the eigenvalue
$\omega^2$ is real due to the Hermiticity (self-adjointness
\cite{bernstein_etal58}) of the linearized force and kinetic energy
operators.  They use finite-element or finite-difference methods to
convert the infinite-dimensional PDE eigenvalue problem to an
approximating finite-dimensional matrix problem.  An alternative
approach is to use local analysis using the ballooning representation
and to attempt semiclassical quantization to estimate the global
spectrum \cite{dewar-glasser83,cuthbert_etal98,redi_etal02}.

In order properly to verify the convergence of these codes  in
three-dimensional geometry it is essential to understand the nature of
the spectrum---if it is quantum-chaotic then convergence of individual
eigenvalues cannot be expected and a statistical description must be
used \cite{gutzwiller90,mehta91,stoeckmann99,haake01}.

This is perhaps of most importance in understanding the spectrum in
three-dimensional magnetic confinement geometries, in particular the
various stellarator experiments currently running or under
construction.  These devices are called three-dimensional because they
possess no continuous geometrical symmetries, and thus there is no
separation of variables to reduce the dimensionality of the eigenvalue
problem.  It has been shown \cite{dewar-cuthbert-ball01} that the
semiclassical limit (a Hamiltonian ray tracing problem) for ballooning
instabilities in such geometries may be strongly chaotic because there
are no ignorable coordinates in the ray Hamiltonian.

However, the present paper discusses the opposite limit, a system with
a sufficient number of symmetries to make the ray Hamiltonian
integrable and the eigenvalue problem separable.  The geometry is the
circular cylinder, periodic in the $z$-direction to make it
topologically toroidal---we shall refer to the $z$-direction as the
toroidal direction and the azimuthal, $\theta$-direction as the poloidal
direction.  The study of this separable system will provide a
baseline for comparison with the three-dimensional toroidal case in
future work.  The overall goal of the paper is to determine if
the ideal-MHD spectrum falls within the same universality class as
that of typical waves in separable geometries or, if not, what might
cause it to differ.

Berry and Tabor \cite{berry-tabor77} show that the distribution
function $P(s)$ for the spacing of adjacent energy levels (suitably
scaled) in a generic separable quantum system with more than one
degree of freedom is $\exp(-s)$, as for a Poisson process with levels
distributed at random.  They also show that the spectrum of uncoupled
quantum oscillators is nongeneric even when the frequency ratios are
not commensurate, in which case $P(s)$ peaks about a nonzero value of
$s$ (as also occurs in nonintegrable, chaotic systems---the ``level
repulsion'' effect).  A more surprising departure from the Poisson
distribution was found by Casati \emph{et al.}
\cite{casati-chirikov-guarneri85} for waves in a rectangular box with
irrational aspect ratio, but the departure was very small.  Level
spacing statistics are discussed also in the standard monographs on
quantum chaos \cite{gutzwiller90,mehta91,stoeckmann99,haake01}.

In contrast with quantum mechanics, where the continuous spectrum
arises from the unboundedness of configuration space, the ideal-MHD
essential spectrum arises from the unboundedness of Fourier
space---there is no minimum wavelength.  This is an unphysical
artifact of the ideal MHD model because, in reality, low-frequency
instabilities with $|\kvec_{\perp}|$ much greater than the ion Larmor
radius, $a_{\mathrm{i}}$, cannot exist (where $\kvec_{\perp}$ is the
projection of the local wavevector into the plane perpendicular to the
magnetic field $\Bvec$).  Indeed, ideal MHD breaks down in various
ways at large $|\kvec_{\perp}|$, with dissipative and drift effects
coming into play.

In this paper we do not attempt to model finite-Larmor-radius
stabilization, but instead simply restrict the poloidal mode spectrum
to $m \leq \mmax$ and study the scaling of the spectrum at large
$\mmax$.  The nature of the dispersion relation is such that the
toroidal mode numbers $n$ relevant to the spectrum are also
restricted.  In a matrix eigenvalue code such as CAS3D or TERPSICHORE
our procedure corresponds to using an arbitrarily fine radial mesh
but truncating the toroidal and poloidal basis set.

The eigenvalue equation for a reduced MHD model of a stellarator is
presented in Sec.~\ref{sec:Model}.  We study a plasma in which the
Suydam criterion \cite{suydam58} for the stability of interchange
modes is violated, so the number of unstable modes is infinite.  

Section~\ref{sec:Spectrum} is devoted to developing an understanding
of the dependence (the \emph{dispersion relation}) of the eigenvalues
on the radial, poloidal and toroidal mode numbers, $l$, $m$, and $n$,
respectively.  As $m$ and $n$ approach infinity, keeping $\mu \equiv
n/m$ fixed, the growth-rate eigenvalues asymptote to a constant, the
Suydam growth rate, depending only on $\mu$ and the radial mode number
$l$.  We use a combination of perturbation expansion in $1/m$ and
numerical solution of the eigenvalue equation using a new
transformation to Schr\"{o}dinger form that is applicable over the
whole range of $m$, from $O(1)$ to $\infty$.  This generalizes the
approach of Cheremhykh and Revenchuk \cite{cheremhykh-revenchuk92},
which was limited to the $m = \infty$ Suydam eigenvalue problem.  We
compare some of the asymptotic results in
\cite{cheremhykh-revenchuk92} with our numerical solutions.  Our
perturbation expansion shows that the correction to the Suydam limit
goes as $1/m^2$.  Contrary to usual experience
\cite{sugama-wakatani89}, our numerical solutions show that the growth
rates do not always approach the Suydam values from below as $m
\rightarrow \infty$.

In Sec.~\ref{sec:S0} we examine the part of the spectrum involving the
most unstable modes, which is essentially two-dimensional because only
the lowest-order radial mode, $l=0$, contributes.  We relate the
considerable amount of structure observed in the spectrum to the Farey
sequences of rational values of the rotational transform (winding
number) of the equilibrium magnetic field.  Low-order rationals have
associated eigenvalue sequences giving a regular distribution of
eigenvalues locally more like the spectrum of a one-dimensional system
than a two-dimensional one.

In Sec.~\ref{sec:Weyl} we derive the analog of the Weyl formula for
the average density of states, including an asymptotic analysis of the 
large-$l$ limit. In Sec.~\ref{sec:sDist} we show
level spacing distributions $P(s)$.  Since we are interested in large
$m$ we first try approximating the eigenvalues by their corresponding
asymptotic Suydam limit.  This gives a very singular distribution with
a delta-function-like spike at the origin
\cite{dewar-nuehrenberg-tatsuno04} due to the extremely degenerate
nature of the spectrum in this approximation.  By contrast the
distribution for the exact spectrum has no spike at the origin,
showing that the small $1/m^2$ corrections break the degeneracy
sufficiently to completely change the statistics.

We examine the statistics for the $l=0$ and $l=1$ spectra, both
individually and combined (in the low-growth-rate region where they
overlap).  We have examined sufficiently large data sets to show
convincingly that the statistical distributions are not Poissonian,
though that of the combined $l=0$ and $l=1$ spectrum is closest.  We
also split the $l=0$ spectrum into two halves to remove overlap of
spectra arising from different parts of the plasma.  These split
spectra exhibit a much more dramatic departure from Poisson
statistics, showing that the ideal-MHD interchange spectrum is indeed
nongeneric in the sense of Berry and Tabor \cite{berry-tabor77}.

\section{Choice of model eigenvalue equation}
\label{sec:Model}

The grand context of this paper is the three-dimensional linearized
ideal MHD problem---to solve, under appropriate boundary conditions,
the equation of motion
\begin{equation}
		\rho\partial_t^2\xivec = \Ften\cdot\xivec
		\label{eq:eqmotion}
\end{equation}
for small displacements $\xivec(\rvec,t)$ of the MHD fluid about a
static equilibrium state, where $\rho(\rvec)$ is the equilibrium mass
density, $\rvec$ is position, $t$ is time, and $\Ften$ is a Hermitian
linearized force operator \cite{bernstein_etal58} under the inner
product $\int d^3 x\,\xivec^{*}\dotv\Ften\dotv\xivec$ and suitable
boundary conditions. (Superscript * denotes complex 
conjugation---we can take $\xivec$ to be complex because all
the coefficients in $\Ften$ are real, so the real and imaginary parts
of $\xivec$ obey the same equation.)  

Most modern magnetic confinement fusion experiments, in particular
tokamaks and stellarators, are toroidal.  Though not guaranteed for
arbitrary three-dimensional systems, the equilibrium magnetic field
$\Bvec(\rvec)$ is normally assumed to be \emph{integrable} in the
sense that all field lines lie on invariant tori (magnetic surfaces)
nested about a single closed field line (the magnetic axis).  Within
each toroidal magnetic surface a natural angular coordinate system is
set up, with the poloidal angle $\theta$ increasing by $2\pi$ for each
circuit around the short way and the toroidal angle $\zeta$ increasing
by $2\pi$ for each circuit the long way.  Each surface is
characterized by a magnetic winding number, the \emph{rotational
transform} $\iotabar$, being the average poloidal rotation of a field
line per toroidal circuit, $\langle d\theta/d\zeta\rangle$, over an
infinite number of circuits.  (In tokamak physics the inverse, $q
\equiv 1/\iotabar$, is normally used as the rotation number.)

In this paper we study an effectively circular-cylindrical MHD
equilibrium, using cylindrical coordinates such that the magnetic axis
coincides with the $z$-axis, made topologically toroidal by periodic
boundary conditions.  Thus $z$ and the toroidal angle $\zeta$ are
related through $\zeta \equiv z/R_0$, where $R_0$ is the major radius
of the toroidal plasma being modeled by this cylinder.  The poloidal
angle $\theta$ is the usual geometric cylindrical angle and the
distance $r$ from the magnetic axis labels the magnetic surfaces (the
equilibrium field being trivially integrable in this case).  The
plasma edge is at $r = a$.

In the cylinder there are two ignorable coordinates, $\theta$ and
$\zeta$, so the components of $\xivec$ are completely factorizable into
products of functions of the independent variables separately.  In
particular, we write the $r$-component as
\begin{equation}
		r\xi_r = \exp (im\theta )\exp (-in\zeta)\varphi(r) \;,
		\label{eq:sep}
\end{equation}
where the periodic boundary conditions quantize $m$ and $n$ to 
integers and we choose to work with the stream function 
$\varphi(r) \equiv r\xi_r(r)$.  

Since the primary motivation of this paper is stellarator physics, we
use the reduced MHD ordering for large-aspect stellarators
\cite{strauss80,wakatani98}, averaging over helical ripple to reduce
to an equivalent cylindrical problem
\cite{kulsrud63,tatsuno-wakatani-ichiguchi99}.  The universality class
should be insensitive to the precise choice of model as long as it
exhibits the behavior typical of MHD instabilities in a cylindrical
plasma, specifically the existence of interchange instabilities and
the occurrence of accumulation points at finite growth rates.

We nondimensionalize by measuring the radius $r$ in units of the minor
radius of the plasma column, $a$, and the time $t$ in units of the
poloidal Alfv\'en time $\tau_{\mathrm{A}} = R_0 \sqrt{\mu_0 \rho} /
B_0$, where $B_0$ is the toroidal magnetic field and $\mu_0$ is the
permeability of free space.  Thus $\omega$ is in units of $\tau_{\rm
A}^{-1}$.  Defining $\lambda \equiv \omega^2$ we seek the spectrum of
$\lambda$-values satisfying the scalar equation
\begin{equation}
	 L \varphi = \lambda M\varphi
		\label{eq:eigvaleqn}
\end{equation}
under the boundary conditions $\varphi(0) = 0$ at the magnetic axis
and $\varphi(1) = 0$, appropriate to a perfectly conducting wall at
the plasma edge. The operators $L$ and $M$ given below 
are Hermitian under the inner product defined, for arbitrary functions 
$f$ and $g$ satisfying the boundary conditions, by
\begin{equation}
		\langle f,g\rangle \equiv \int_0^1 f^{*}(r)g(r)\;r\,dr \;.
		\label{eq:inner}
\end{equation}
The weight factor $r$ in the inner product is a Jacobian factor coming
from $d^3x = rdrd\theta dz$.

The operator $L$ is given by
\begin{eqnarray}
	 L & \equiv & -\frac{1}{r}\frac{d}{dr}(n - m\iotabar)^2 r\frac{d}{dr} 
	 \nonumber \\  & & \mbox{}
	 +\frac{m^2}{r^2}
	 	\left[
		(n - m\iotabar)^2 - \Ds  + 
         \frac{\iotaddot}{m}(n-m\iotabar)
	 	\right] \;,
 	 \label{eq:Ldef}
\end{eqnarray}
where the Suydam stability parameter $\Ds$ is
\begin{equation}
	 \Ds \equiv -\frac{\beta_0}{2\epsilon^2}p'(r)\Omega'(r) \;,
	 \label{eq:Dsdef}
\end{equation}
with $\epsilon \equiv a/R_0 \ll 1$ the inverse aspect ratio, $p(r)$
the plasma pressure normalized to unity at $r=0$, $\beta_0 \equiv
2\mu_0 p_0/B_0^2$  the ratio of plasma pressure to magnetic 
pressure at the magnetic axis, and $\Omega'$ the average field line 
curvature. Here
\begin{equation}
		\Omega \equiv \epsilon^2 N\left(r^2\iotabar  + 2\int r\iotabar dr \right)
		\label{eq:helcurv}
\end{equation}
where the rotational transform is produced by helical current windings
making $N \gg 1$ turns as $\zeta$ goes from $0$ to $2\pi$,
$\Omega'(r)$ giving the averaged field-line curvature.  (Note that
$\epsilon$ cancels out in $\Ds$.)  We use the notation $\dot{f} \equiv
rf'(r)$ for an arbitrary function $f$, so $\iotadot \equiv rd\iotabar/dr$
is a measure of the magnetic shear and $\iotaddot$ measures the
variation of the shear with radius.  The term $\Omega$ is a measure of
the ``magnetic hill'' \cite{wakatani98} that allows pressure energy to
be released by interchanging field lines, thus driving the
interchange instability.

The operator arising from the inertial term in \eq{eq:eqmotion},
\begin{equation}
	 M \equiv  -\nabla_{\perp}^2 = -\frac{1}{r}\frac{d}{dr}r\frac{d}{dr}  + 
	 \frac{m^2}{r^2} \;,
	 \label{eq:Mdef}
\end{equation}
is easily seen to be positive definite under the inner product \eq{eq:inner}.

We observe some differences between \eq{eq:eigvaleqn} and the standard
quantum mechanical eigenvalue problem $H\psi = E\psi$.  One is of
course the physical interpretation of the eigenvalue---in quantum
mechanics the eigenvalue $E \equiv \hbar \omega$ is linear in the
frequency because the Schr\"odinger equation is first order in time,
whereas our eigenvalue $\lambda$ is quadratic in the frequency because
it derives from a classical equation of motion. 

Another difference is that \eq{eq:eigvaleqn} is a \emph{generalized}
eigenvalue equation because $M$ is not the identity operator.  This is
one reason why it is necessary to treat the MHD spectrum explicitly
rather than simply assume it is in the same universality class as
standard quantum mechanical systems.

Just as in ordinary eigenvalue problems the eigenvalue spectrum for
the generalized eigenvalue problem is real, and the eigenfunctions
$\varphi_i$ have a generalized orthogonality property
\begin{equation}
		\langle \varphi_i,M\varphi_j\rangle = \delta_{i,j} \;,
		\label{eq:orthonorm}
\end{equation}
where the normalization has been chosen to make the coefficient of the
Kronecker $\delta$ unity.  Here $i$ and $j$ denote members of the set
$\{l,m,n\}$, where $l$ is the radial node number and the poloidal and
toroidal mode numbers $m$ and $n$, respectively, are defined in
\eq{eq:sep}.  The negative part of the spectrum, $\lambda = -\gamma^2
< 0$, corresponds to instabilities growing exponentially with growth
rate $\gamma$.

Equation~(\ref{eq:eigvaleqn}) is very similar to the normal mode
equation analyzed in the early work on the interchange growth rate in
stellarators by Kulsrud \cite{kulsrud63}.  However, unlike this and
most other MHD studies we are concerned not with finding the highest
growth rate, but in characterizing the complete set of
unstable eigenvalues.

\section{Interchange spectrum}\label{sec:Spectrum}

In this section we discuss the standard unregularized ideal MHD
spectrum.  It is well known that for $\lambda>0$ the spectrum consists
of the Alfv\'en continuum (the slow-magnetosonic continuum being
removed in reduced MHD \cite{mcmillan-dewar-storer04}).  On the
unstable side of the spectrum, $\lambda < 0$, it is also known that
there is an infinity of eigenvalues provided the Suydam interchange
instability criterion \cite{suydam58}
\begin{equation}
    G \equiv \frac{\Ds}{\iotadot^2} > \frac{1}{4}
    \label{eq:Suydam}
\end{equation}
is satisfied over some range of $r$ in the interval $(0,1)$, but the 
details of the spectrum do not appear to have been published before. 

\begin{figure}[tbp]
		\begin{tabular}{cc}
				\includegraphics[scale=0.5]{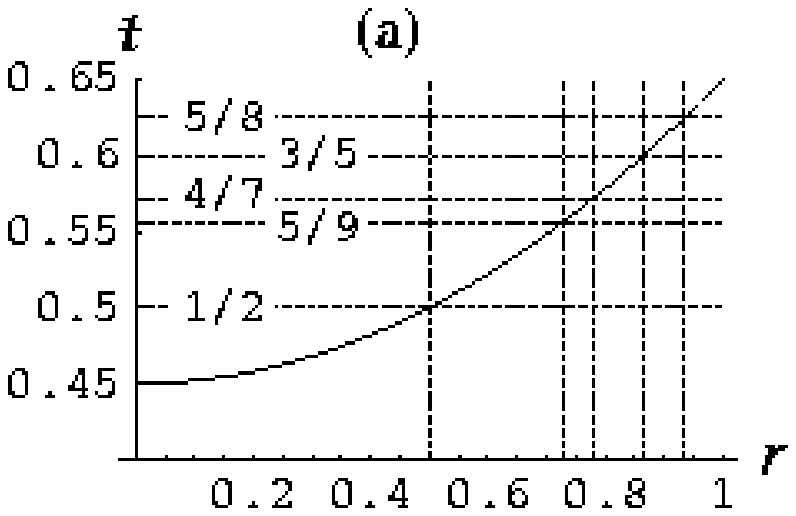} &
				\includegraphics[scale=0.5]{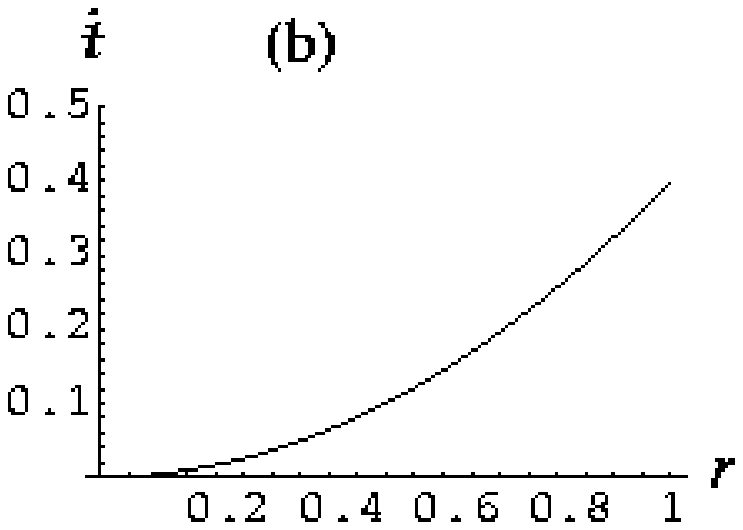} 
		\end{tabular}
		\caption{(a) The rotational transform $\iotabar(r)\equiv 1/q(r)$
		defined by \eq{eq:iotaprof} with $\iotabar_0=0.45$,
		$\iotabar_2=0.2$ (as for all subsequent plots); (b) the magnetic
		shear parameter $\iotadot\equiv r\iotabar'(r)$.  In (a), all
		distinct rational magnetic surfaces $\iotabar = \mu \equiv n/m$
		are shown for $m$ up to 10.}
		\label{fig:iprofiles}
\end{figure}

\subsection{Profiles}
\label{sec:RatSurf}

Interchange instabilities occur only for values of $m$ and $n$ such
that $n - m\iotabar$ vanishes (or at least can be made very small
\cite{tatsuno-wakatani-ichiguchi99}) and therefore it is important to know 
something about the function $\iotabar(r)$. The typical profile of
$\iotabar(r)$ in a stellarator is monotonically increasing in the
interval $[0,a]$ and we shall assume this to be the case here (though
it is not always true in modern stellarators). For the numerical work 
in this paper we use a parabolic profile
\begin{equation}
		\iotabar = \iotabar_0 +  \iotabar_2 r^2
		\label{eq:iotaprof}
\end{equation}
as illustrated in \fig{fig:iprofiles}(a).

Given a rational fraction $\mu = n_{\mu}/m_{\mu}$ in the
interval $[\iotabar(0),\iotabar(a)]$ (where $n_{\mu} $and $m_{\mu}$
are mutually prime) there is a unique radius $\rmu$ such that
\begin{math}%\label{eq:rmudef}
    \iotabar(\rmu) = \mu \;.
\end{math}
Any pair of integers $ (m,n)_{\mu,\nu} \equiv (\nu m_{\mu}, \nu
n_{\mu})$, $\nu = 1, 2, 3, \ldots$ satisfies the resonance condition
\begin{equation}
    n_{\mu,\nu}  - m_{\mu,\nu} \iotabar(\rmu) = 0 \;.
    \label{eq:rescond}
\end{equation}
For example, the set of rationals with $1 < m \leq 10$ in the interval
of $\iotabar$ shown in \fig{fig:iprofiles}(a) is $\{\mu\} = \{
1/2,\;5/9,\;4/7,\;3/5,\;5/8 \}$, as shown in the figure.

\begin{figure}[tbp] 
		\begin{tabular}{cc}
				\includegraphics[scale=0.5]{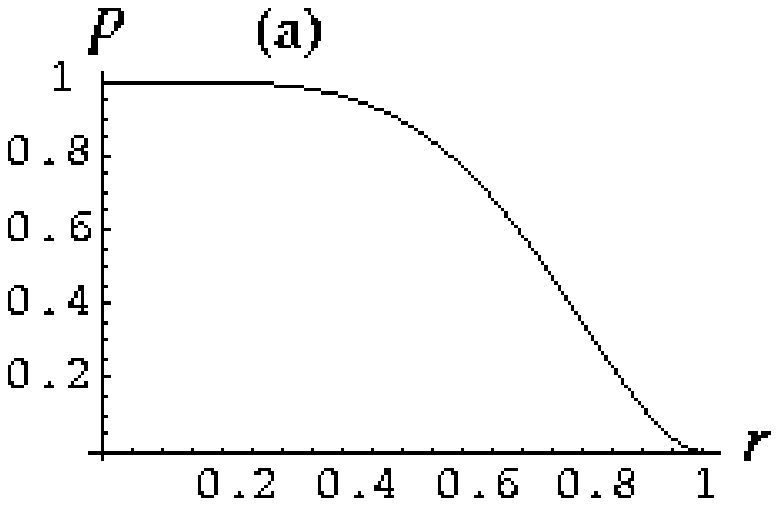} &
				\includegraphics[scale=0.5]{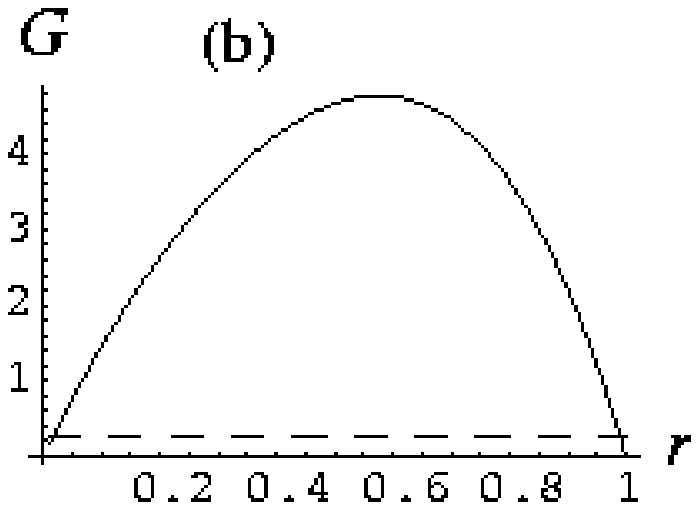} 
		\end{tabular}
		\caption{(a) The pressure profile $p(r)$, \eq{eq:pprof}, used in
		this paper and (b) the Suydam criterion parameter $G(r)$, defined
		in \eq{eq:Suydam} (solid line), and the instability threshold
		$1/4$ (dashed line), showing nearly all the plasma is interchange
		unstable.}
		\label{fig:pprofiles}
\end{figure}

To understand the global spectrum we also need to know something about
the pressure profile.  In this paper we use a broad pressure profile
that is sufficiently flat near the magnetic axis that the Suydam
instability parameter $G$ defined in \eq{eq:Suydam} goes to zero at
the magnetic axis, and for which $p'$ vanishes at the plasma edge
\begin{equation}
		p(r) = 1 - 6\,r^5 + 5\,r^6 \;.
		\label{eq:pprof}
\end{equation}
This profile is shown in \fig{fig:pprofiles}(a) and the resulting 
$G$-profile in \fig{fig:pprofiles}(b).

\subsection{High $m$ and $n$}
\label{sec:Hi_m}

In this subsection we choose a particular rational surface $\rmu$
and restrict attention to pairs $(m,n)$ from the set $\{(m,n)_{\mu,\nu}|\nu =
1,2,3,\ldots\}$ satisfying the condition \eq{eq:rescond}.

Defining a scaled radial variable $x \equiv m(r-\rmu)/\rmu$, we expand all
quantities in inverse powers of $m$, 
\begin{eqnarray}
    L & \equiv & \frac{m^2}{\rmu^2}(L^{(0)} + m^{-1}L^{(1)} + m^{-2}L^{(2)} + \ldots )
    \;, \nonumber \\
    M & \equiv & \frac{m^2}{\rmu^2}(M^{(0)}+ m^{-1}M^{(1)} + m^{-2}M^{(2)} + \ldots ) 
     \label{eq:Expansion}
\end{eqnarray}
Also, $\lambda = \lambda^{(0)} + m^{-1}\lambda^{(1)} +
m^{-2}\lambda^{(2)} $, and similarly for $\varphi$.  The detailed
expressions are given in Appendix~\ref{sec:mm2app} .

We then solve  \eq{eq:eigvaleqn} by equating
the LHS to zero order by order.  At $O(m^0)$, as found by Kulsrud
\cite{kulsrud63}, we have the generalized eigenvalue equation
\begin{equation}
		\mathcal{L}^{(0)}\varphi^{(0)} = 0 \;,
		\label{eq:eigvaleqn0}
\end{equation}
where
\begin{eqnarray}
	\mathcal{L}^{(0)}  & \equiv & L^{(0)} - \lambda^{(0)}M^{(0)} 
	\nonumber \\      & = &
     -\frac{d}{dx}(\iotadot^2 x^2 - \lambda^{(0)}) \frac{d}{dx}  
     + \iotadot^2 x^2 - \lambda^{(0)} - \Ds 
    \label{eq:calL0def}
\end{eqnarray}
with $\iotadot$ and $\Ds$ evaluated at $\rmu$.  For
$\lambda^{(0)} < 0$, \eq{eq:eigvaleqn0} can be solved to give a
square-integrable eigenfunction under the boundary conditions
$\varphi^{(0)} \rightarrow 0$ as $r \rightarrow \pm\infty$ when
$\lambda^{(0)}$ is one of the eigenvalues $\lambda_{\mu,l}$, $l =
0,1,2,\ldots$, denoting the number of radial nodes of the
eigenfunction $\varphi^{(0)} = \varphi_{\mu,l}$.  Note that
$\lambda_{\mu,l}$ depends only on $\mu = n/m$ and is otherwise
independent of the magnitude of $m$ and $n$.  We assume that the
$\varphi_{\mu,l}(r)$, when combined with the continuum generalized
eigenfunctions for $\lambda^{(0)} > 0$, form a complete set.

\subsubsection{Suydam approximation}
\label{sec:CR}

\begin{figure}[btp]
		\begin{tabular}{cc}
				\includegraphics[scale=0.5]{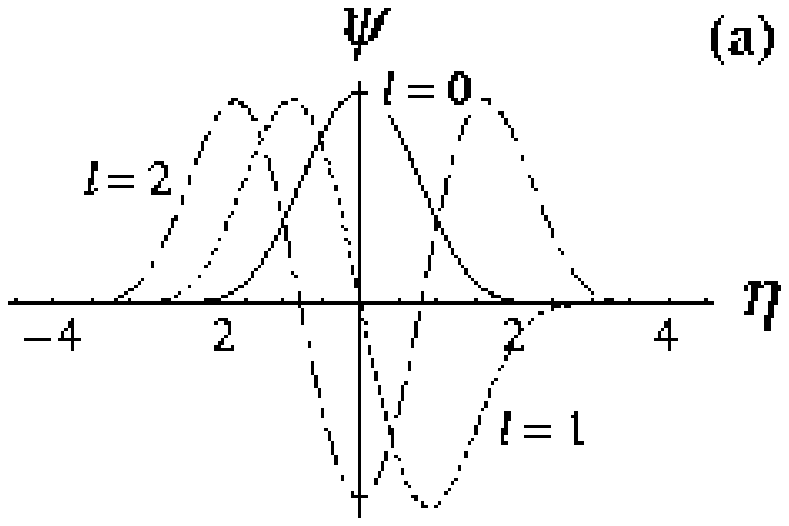} &
				\includegraphics[scale=0.5]{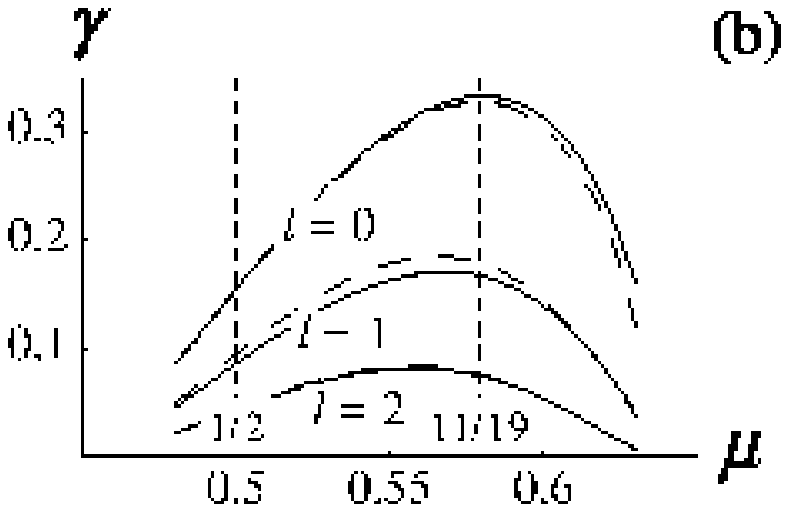} 
		\end{tabular}
		\caption{(a) $m=\infty$ eigenfunctions for the $l = 0$ (solid
		line), $l = 1$ (short dashes) and $l = 2$ (short and long dashes)
		modes at $\mu = 1/2$, arbitrary normalization.  (b) Growth rates
		$\gamma$ \emph{vs.} resonant $\iotabar \equiv \mu$.  Dashed
		lines show approximations \eq{eq:CR4.7} (for $l = 0$) and
		\eq{eq:CR4.5} (for $l = 1$ and $2$).} \label{fig:infm}
\end{figure}

The leading term in the expansion of the eigenvalue in $1/m$ gives the
growth rate in the limit $m \rightarrow \infty$, known as the
\emph{Suydam approximation}.  Restricting attention to unstable modes,
so that $\gamma \equiv (-\lambda)^{1/2}$ is real, we transform
\eq{eq:eigvaleqn0} to the Schr\"odinger form
\cite{cheremhykh-revenchuk92}
\begin{equation}
		 \frac{d^2\psi}{d\eta^2} + Q(\eta)\psi = 0 \;,
		\label{eq:Schrodinger}
\end{equation}
where
\begin{equation}
		Q=Q_0(\eta|\gamma,\mu) \equiv  G - \quarter - \quarter\sech^2\,\eta
		                       - \Gamma^2\cosh^2\eta  \;,
		\label{eq:Q0}
\end{equation}
with $G \equiv G(r_{\mu})$ defined as in \eq{eq:Suydam}, $\Gamma
\equiv \gamma/\iotadot(r_{\mu})$, $\eta$ defined through $x \equiv
\gamma\sinh\eta/\iotadot(r_{\mu})$, and $\psi \equiv
(\cosh\eta)^{1/2}\varphi(x)$.  [In
Ref.~\onlinecite{cheremhykh-revenchuk92} \eq{eq:Q0} is derived from
the Fourier transform of \eq{eq:eigvaleqn0}, but we can also use the
real-space version as the equation shares with the quantum oscillator
the remarkable property of having the same general form in both
Fourier space and real space.]

Cheremhykh and Revenchuk \cite{cheremhykh-revenchuk92} (CR) have made
an extensive study of the eigenvalues of \eq{eq:Schrodinger} using the
semiclassical quantization condition
\begin{equation}
		\oint Q_0(\eta)^{1/2}\,d\eta = (2l +1)\pi  \;.
		\label{eq:Quant0}
\end{equation}
which follows from the WKB ansatz $\psi = A(\eta)\exp \pm i\int
Q_0^{1/2}\,d\eta$.  CR derive several approximations, useful in
appropriate limits, improving on the earlier result of Kulsrud
\cite{kulsrud63}.  In this paper we use two of their results
to compare with numerical solutions of \eq{eq:Schrodinger}. The first 
is Eq.~(4.5) of \cite{cheremhykh-revenchuk92}
\begin{equation}
		\Gamma \approx \frac{4\sigma}{e}\exp
		  \left[
				-\frac{(l+\half)\pi}{2\sigma}
				-\frac{1}{4\sigma^2}
			\right]	\;	,
		\label{eq:CR4.5}
\end{equation}
which [combining the criteria given in CR's Eqs.~(4.4) and
(4.12)]  is applicable when $\sigma \equiv (G-1/4)^{1/2} \gg 1/2$. 
The second CR result we use is their Eq.~(4.7)
\begin{equation}
		\Gamma^2 \approx \frac{G - (2l+1)G^{1/2}}{1 + (4G)^{-1}} \;,
		\label{eq:CR4.7}
\end{equation}
applicable when $G \gtrsim \Gamma^2 \gg 1$.

As is seen from \fig{fig:infm}, \eq{eq:CR4.7} gives a remarkably good
approximation to the growth rate of the most unstable radial
eigenmode, $l=0$, and \eq{eq:CR4.5} gives a good approximation for the
higher-$l$ modes (the semiclassical quantization being strictly
justifiable only for large $l$).  The growth-rate maxima for each $l$
occur close to the maximum of $G$ (and hence $\Gamma)$, but not
exactly owing to the $\iotadot$ factor in the definition $\Gamma \equiv
\gamma/\iotadot(r_{\mu})$.

From \eq{eq:CR4.7} we see that, provided the Suydam criterion $G >
1/4$ is satisfied, there is an infinity of growth rate eigenvalues
accumulating exponentially toward the origin from above (so the
$\lambda$-values accumulate from below) in the limit $l \rightarrow
\infty$.

Perhaps less widely appreciated (because $m$ and $n$ are normally
taken to be fixed) is the fact that there is also a point of
accumulation of the eigenvalues of \eq{eq:eigvaleqn} at each
$\lambda_{\mu,l}$ as $\mmax \rightarrow \infty$ with $l$ fixed.  To
break the degeneracy of $\lambda^{(0)}$ we must proceed further with
the expansion in $1/m$.

\subsubsection{$1/m^2$ corrections}
\label{sec:mm2txt}

Proceeding with the expansion \eq{eq:Expansion}, the calculation goes
through much as in standard time-independent quantum perturbation
theory \cite[e.g.]{landau-lifshitzQM}. 

The lowest order eigenvalues and eigenfunctions are, as found in
Sec.~\ref{sec:Hi_m}, $\lambda^{(0)} = \lambda_{\mu,l}$ and
$\varphi^{(0)} = \varphi_{\mu,l}(x)$, respectively.  The $O(1/m)$
correction, $\lambda^{(1)}$, vanishes identically from parity
considerations---$\varphi_{\mu,l}(x)$ is either an even or odd
function so its contribution to the matrix elements of $L^{(1)}$ and
$M^{(1)}$ between $\varphi^{(0)}$ and $\varphi^{(0)}$ is even.  On the
other hand, $L^{(1)}$ and $M^{(1)}$ are odd, so $\lambda^{(1)} \equiv
0$.  (This contrasts with the finite-aspect-ratio toroidal case where
toroidal coupling of Fourier harmonics of different $m$ to form
ballooning modes leads to a nonvanishing $1/n$ correction
\cite{connor-hastie-taylor79,dewar-chance-glasser-greene-frieman79}.)

The first nonvanishing correction term is thus
\begin{eqnarray}
		\lambda^{(2)} & = &
		\langle \mu,l |\mathcal{L}^{(2)}| \mu,l \rangle
 		\nonumber \\  & & \mbox{}
		-\sum_{l' \neq l} 
		   \frac{
			 \langle \mu,l |\mathcal{L}^{(1)}| \mu,l' \rangle
			 \langle \mu,l' |\mathcal{L}^{(1)}| \mu,l \rangle
			 }
			 {\lambda_{\mu,l'} - \lambda_{\mu,l}} \;,
		\label{eq:lam2}
\end{eqnarray}
where the sum over $l'$ is taken to include an integration over the
continuum.  The operators $\mathcal{L}^{(i)}\equiv L^{(i)} -
\lambda_{\mu,l}M^{(i)}$ are the higher-order generalizations of
$\mathcal{L}^{(0)}$, defined by \eq{eq:calL0def}.  The $m = \infty$
matrix elements of any operator $\mathcal{F}$ are defined by
\begin{equation}
		\langle \mu,l' |\mathcal{F}| \mu,l'' \rangle
		\equiv 
		\int_{-\infty}^{\infty}
		\varphi_{\mu,l'}^{*}(x)\mathcal{F}\varphi_{\mu,l''}(x) \;dx \;,
    \label{eq:inner0}
\end{equation}
with the eigenfunctions $\varphi_{\mu,l}(x)$ being normalized so that
$\langle \mu,l |M^{(0)}| \mu,l \rangle = 1$.  Note that, with the
operators $L$ and $M$ defined as in Eqs.~(\ref{eq:Ldef}) and
(\ref{eq:Mdef}) , $\mathcal{L}^{(i)}$ is Hermitian under the inner
product used in \eq{eq:inner0} only for $i = 0$.  However, it can be
made Hermitian at arbitrary order by the redefinitions $L \mapsto rL$
and $M \mapsto rM$, which puts the eigenvalue equation into
Sturm--Liouville form.

\begin{figure}[tbp]
		\begin{tabular}{cc}
				\includegraphics[scale=0.5]{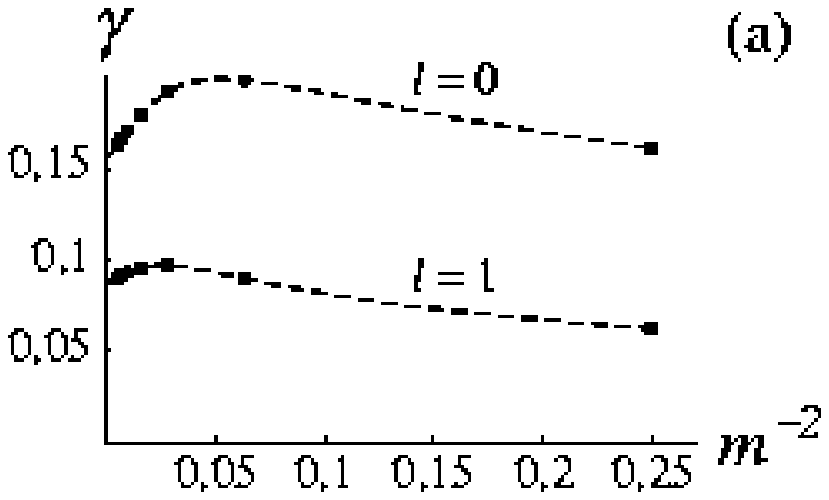} &
				\includegraphics[scale=0.5]{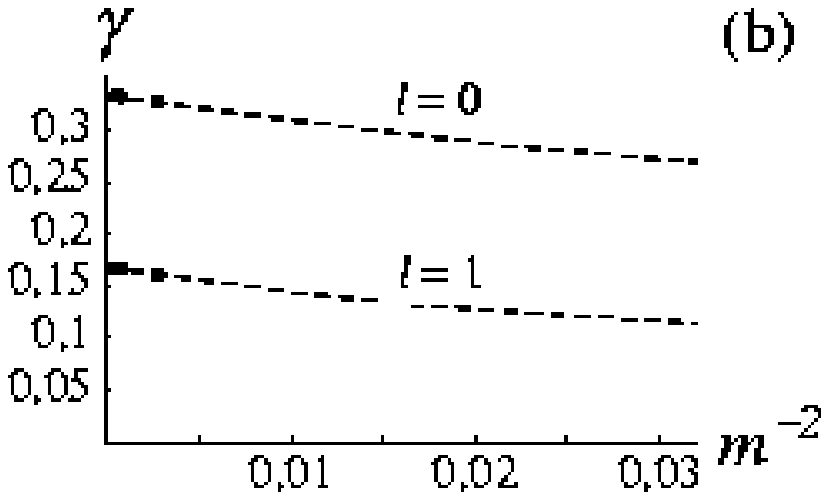} 
		\end{tabular}
		\caption{Growth rates $\gamma_{l,m,n} \equiv (-\lambda_{l,m,n})^{1/2}$
		\emph{vs.}  $m^{-2}$ for $l=0$ and $1$, found by numerical solution of
		\eq{eq:eigvaleqn} (Sec.~\ref{sec:finm}): (a) $n/m = 1/2$ ($m =
		2,4,6,\ldots$), and (b) $n/m = 11/19$ ($m = 19,38,57,\ldots$).  At
		high $m$ the dependence becomes linear, in qualitative
		agreement with Sec.~\ref{sec:mm2txt}.}
		\label{fig:finm}
\end{figure}

\begin{figure}[tbp]
		\begin{tabular}{cc}
				\includegraphics[scale=0.5]{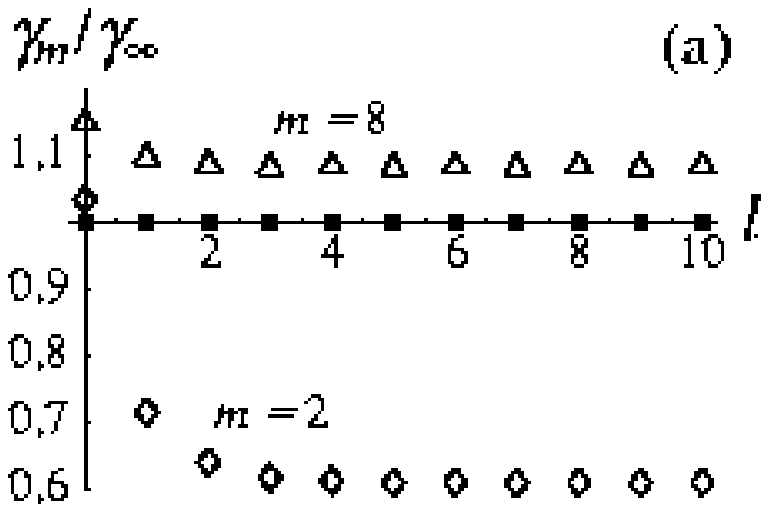} &
				\includegraphics[scale=0.5]{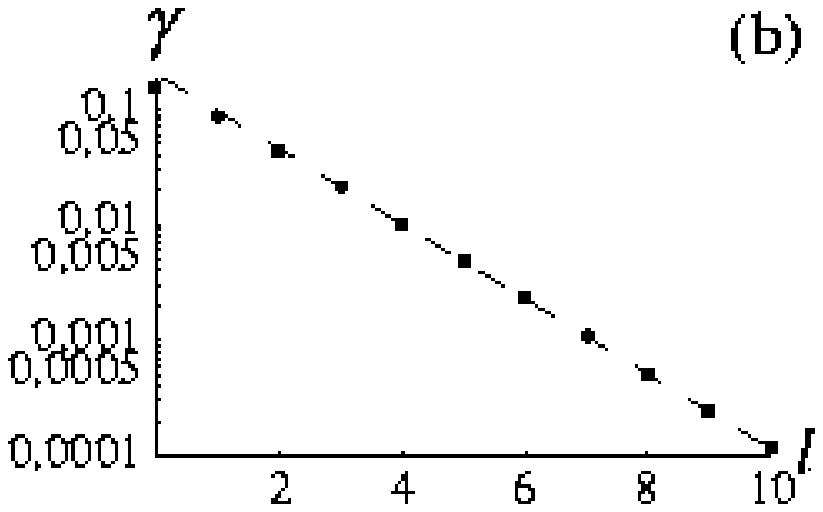} 
		\end{tabular}
		\caption{Growth rates \emph{vs.}  radial node number $l$: (a) found by
		numerical solution of \eq{eq:eigvaleqn} (Sec.~\ref{sec:finm}) for
		$(m,n) = (2,1)$ (diamonds) and $(8,4)$ (triangles), normalized to
		the $\mu = 1/2$, infinite-$m$ results (filled boxes); (b) $\mu =
		1/2$, infinite-$m$ results (points) and asymptotic result
		\eq{eq:CR4.5} (dashed line).}
		\label{fig:lScan}
\end{figure}

As in quantum mechanics \cite[e.g.]{landau-lifshitzQM}, if
$\mathcal{L}^{(1)}$ is Hermitian the contribution of the second term
on the right hand side of \eq{eq:lam2} is always negative for the
lowest eigenvalue, $\lambda^{(0)}_0$, because $\lambda_{l'}^{(0)} -
\lambda_l^{(0)} > 0$.  However, in ideal MHD a positive contribution
from the first term usually dominates and the infinite-$m$ mode is
most unstable \cite{sugama-wakatani89}.  As seen in \fig{fig:finm},
this is not always the case: $\lambda^{(2)}_{l=0}$ is negative for
$\mu=1/2$, but positive for $\mu = 11/19 = 0.578947\ldots$.

The latter value of $\mu$ is very close to the value giving the global
maximum Suydam growth rate (see \fig{fig:infm}).  Thus, in the special
case studied here and in accordance with conventional wisdom, the
\emph{global} maximum interchange growth rate occurs at $m = \infty$.
Both these results are intuitively reasonable---the eigenfunctions
become increasingly localized as $m \rightarrow \infty$, so the
highest growth rate is obtained by localizing in the ``most unstable''
region of the plasma, where $\iotabar \approx 11/19$.  On the other
hand, modes which localize in ``less unstable'' regions as $m
\rightarrow \infty$ can achieve a higher growth rate at finite values
of $m$ because their more extended finite-$m$ eigenfunctions overlap
the more unstable region and tap into the free energy from the
pressure gradient in this region.

Since the eigenvalues approach $\lambda_{\mu,l}$ as $1/m^2$, there is
an infinity of modes in the neighborhood of each $\lambda_{\mu,l}$ in
the limit $\mmax \rightarrow \infty$.  That is, they are
finite-growth-rate accumulation points of the complete spectrum.
Because the rationals $\mu$ are dense on the interval
$(\iotabar(r_1),\iotabar(r_2))$, where $(r_1,r_2)$ is the region in
which the Suydam instability criterion is satisfied, and because
$\lambda_{\mu,l}$ in general depends continuously on $\mu$, the
accumulation points $\lambda_{\mu,l}$ fill the interval
$(-\gamma_{\mathrm{max}}^2,0)$ densely.  This is the part of the
unstable spectrum called the ``accumulation continuum'' by Spies and
Tataronis \cite{spies-tataronis03}, though ``accumulation essential
spectrum'' might be better terminology mathematically.

\subsection{Finite $m$ and $n$}
\label{sec:finm}

In order to calculate arbitrarily high or low-$m$ eigenfunctions we
generalize the transformation in Sec.~\ref{sec:CR} by the change of
variable from $r$ to a new independent variable $\eta$ such that
\begin{equation}
		m\iotabar(r) - n  \equiv \gamma \sinh \eta \;,
		\label{eq:etadef}
\end{equation}
and a new dependent variable $\psi(\eta)$ such that
\begin{equation}
		\varphi  \equiv (\iotadot\cosh\eta)^{-1/2}\psi(\eta) \;,
		\label{eq:psidef}
\end{equation}
so that \eq{eq:eigvaleqn} becomes the Schr\"odinger equation
\eq{eq:Schrodinger}, but with $Q_0$ replaced by
\begin{eqnarray}
		Q & \equiv  & G(\eta) - \frac{1}{4} - \frac{\sech^2\eta}{4}
							- \frac{\gamma^2}{\iotadot^2}\cosh^2\eta  
		\nonumber \\ & & \mbox{}
		+ \frac{\tanh\eta}{2\iotadot}\frac{d\iotadot}{d\eta}
		- \frac{1}{2\iotadot}	\frac{d^2\iotadot}{d\eta^2}
		+ \frac{1}{4\iotadot^2}\left(\frac{d\iotadot}{d\eta}\right)^2	\;.
		\label{eq:Q}
\end{eqnarray}
Where $\iotadot\equiv rd\iotabar/dr$ is as defined in previous sections,
but expressed in terms of $\eta$.

Differentiating \eq{eq:etadef} we find
\begin{equation}
		\frac{dr}{d\eta} = \frac{\gamma \cosh\eta}{\iotadot}\,\frac{r}{m} \;.
		\label{eq:drdeta}
\end{equation}
Thus, in the large-$m$ limit, equilibrium parameters such as $G$ and
$\iotadot$ are slowly varying functions of $\eta$, e.g.
$d\iotadot/d\eta = O(1/m)$ and $d^2\iotadot/d\eta^2 = O(1/m^2)$. 
Comparing \eq{eq:Q} with \eq{eq:Q0} we see that, to leading order in
$1/m$, $Q = Q_0$ but with $\iotadot$ now a slow variable rather than a
strict constant.

With the simple form for $\iotabar$, \eq{eq:iotaprof}, assumed in this
paper, \eq{eq:etadef} is easily inverted to give $r(\eta)$, and also a
cancellation occurs between the terms
$\iotadot'(\eta)\tanh\eta/2\iotadot$ and
$\iotadot''(\eta)/2\iotadot$, so that the exact $Q$ is not much more
complicated than $Q_0$.  The eigenvalues in Figs.~\ref{fig:finm} and
\ref{fig:lScan} were computed by integrating \eq{eq:Schrodinger} with
$Q_0$ replaced by the exact $Q$ and with the appropriate finite
boundary conditions.  Low-$m$ results were checked against those from
an untransformed shooting code.  The dashed lines represent the
results of scans through unquantized, noninteger values of $m$ to show
the smooth, but not necessarily monotone, functional dependence of
$\gamma$ on $m$

\section{$l=0$ spectrum}
\label{sec:S0}

\begin{figure}[tbp]
		\begin{tabular}{cc}
				\includegraphics[scale=0.45]{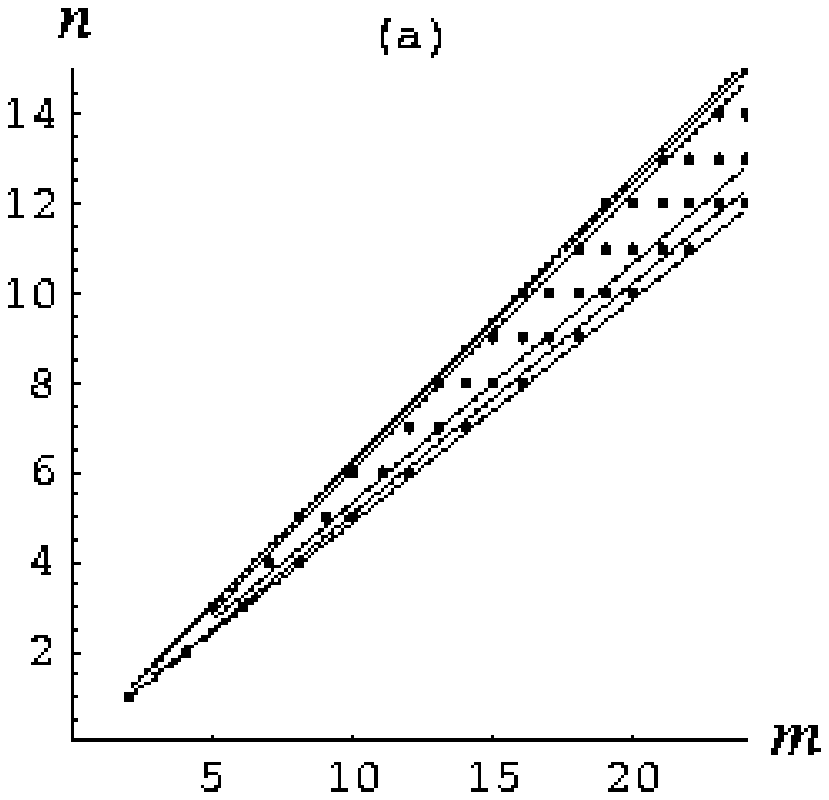} &
				\includegraphics[scale=0.45]{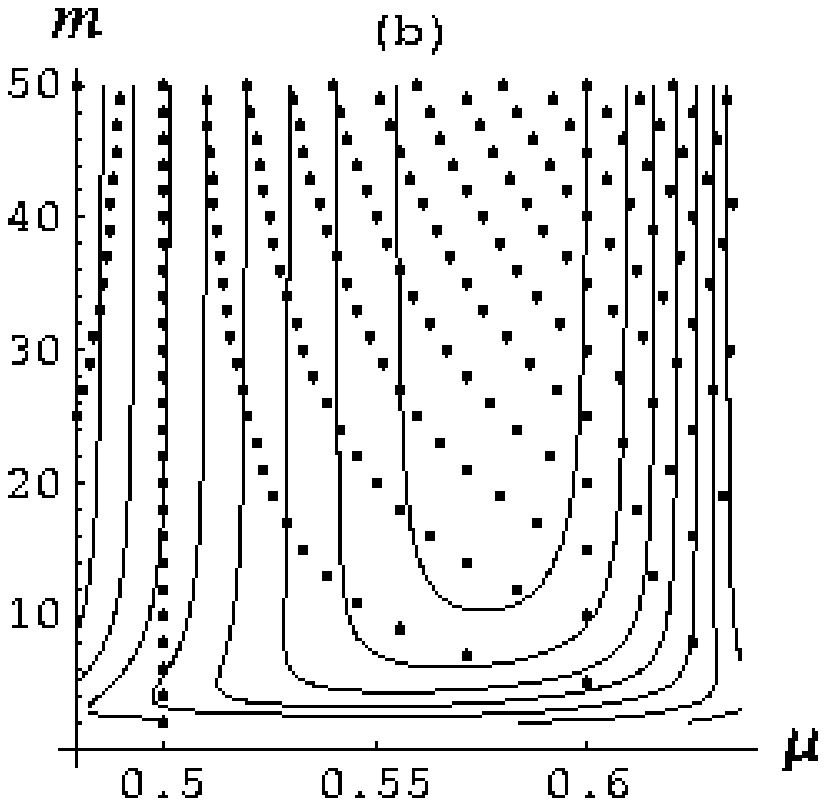} 
		\end{tabular}
		\caption{(a) Lattice of quantum numbers on which the part of the
		spectrum between the ``ground state'' and the threshold for the
		entry of the $l=1$ mode is defined, and the unbounded contours of
		constant eigenvalue.  (b) The same, in $\mu \equiv n/m$, $m$
		space.}
		\label{fig:mnlattice}
\end{figure}

The most unstable modes are those with radial node number $l=0$.
Thus we first consider the set $S_0 \equiv \{\lambda_{0,m,n}|1 \leq m
\leq \mmax, m\mu_{\rm min} < n < m\mu_{\rm max}\}$, where $m$ and $n$
are integers and $\mu_{\rm min}$ and $\mu_{\rm max}$ are chosen to give
the desired range of $\gamma$.  As we shall be rescaling the
eigenvalues prior to statistical analysis, it makes no difference
whether we work with the spectrum of growth rates $\gamma$ or the
eigenvalues $\lambda \equiv -\gamma^2$.  However the latter choice
makes it clearer that the analog of the quantum-mechanical ground state
is the most rapidly growing mode---denoting the maximum growth rate
of the $l=0$ mode by $\gamma_0$, the minimum $\lambda$ is $\lambda_0
= -\gamma_{\rm max}^2= -\gamma_0^2$.

The spectrum is defined on the fan-like subset of the two-dimensional
quantum-number lattice depicted in \fig{fig:mnlattice}(a).  Also shown
are contours of constant $\gamma$ (or $\lambda$), regarded as a
continuous function of $m$ and $n$, which are seen more clearly in
\fig{fig:mnlattice}(b).  Here we see a striking contrast with more
generic systems \cite{berry-tabor77}, where the constant-eigenvalue
contours are segments of topological circles enclosing the origin.
In the ideal-MHD case the contours are topologically hyperbolic, with
asymptotes radiating from the origin toward infinity.

\begin{figure}[tbp]
		\includegraphics[scale=0.6]{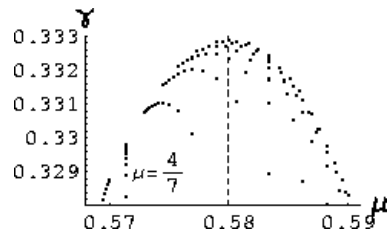} 
		\caption{Growth-rate eigenvalues of $l=0$ modes near the maximum
		growth rate \emph{vs.} $\mu \equiv n/m$.  The ensemble shown is for
		$\mmax = 100$.}
		\label{fig:gamvsmu}
\end{figure}

An interesting representation of the $l=0$ spectrum is shown in
\fig{fig:gamvsmu}.  A great deal of structure can be discerned,
determined by the number-theoretic properties of the interval of $\mu$
depicted.  For instance, focusing on the low-order rational number
$4/7$ we define spectral subsets $S_0(N/M|4/7) \equiv
\{\lambda_{0,m,n}| m = M+7k, n = N+4k, k =
0,1,2,\ldots,[(\mmax-N)/7]\}$, where $[x]$ denotes the largest integer
$\leq x$.

These spectral sequences all accumulate toward the same Suydam
eigenvalue $\lambda_{4/7,0}$ as $\mmax \rightarrow \infty$
independently of the choice of $M$ and $N$.  However the rapidity of
this approach is sensitive to the choice of $M/N$.  For instance we
see in \fig{fig:gamvsmu} the most rapidly converging sequence,
$S_0(4/7|4/7)$, as a set of points accumulating vertically from below
toward the Suydam eigenvalue.  Other sequences on either side of
$S_0(4/7|4/7)$ approach the accumulation point obliquely and much more
slowly---for $\mmax=100$ they visibly have some distance to go.  The
sequence immediately to the left of $S_0(4/7|4/7)$ is $S_0(1/2|4/7)$,
while that to the right is $S_0(3/5|4/7)$, 1/2 and 3/5 being the
immediate neighbors of 4/7 in the Farey sequence \cite[p.
300]{niven-zuckerman-montgomery91} of order 7 (the first order at
which $4/7$ appears), with the $\mu$-values corresponding to
$S_0(1/2|4/7)$ and $S_0(3/5|4/7)$ providing the immediate neighbors of
$4/7$ in each higher-order Farey sequence.

In discussing the structure of the spectrum it is useful to partition
$S_0$ into two subsets, $S^{-}_0$ and $S^{+}_0$, according as the
points are  to the left or right, respectively, of the dashed vertical
line shown in \fig{fig:gamvsmu} passing through the point of maximum
growth rate.

The sequences $S_0(1/2|4/7)$ and $S_0(3/5|4/7)$ accumulate toward
$\lambda_{4/7,0}$, but slower [$O(1/\mmax)$] than does $S_0(4/7|4/7)$
[$O(1/\mmax^2)$, from Sec.~\ref{sec:mm2txt}].  Thus there is a gap
containing $\lambda_{4/7,0}$ within which $S_0(4/7|4/7)$ contributes
$O(\mmax)$ points to $S^{-}_0$, while other sequences contribute at 
most a set of $O(1)$ points.

Within the gap the spectrum $S^{-}_0$ is essentially one-dimensional,
being indexed by the single quantum number $k$.  In the full spectrum,
$S_0 = S^{-}_0 \cup S^{+}_0$, $O(\mmax)$ unrelated eigenvalues from
$S^{+}_0$ appear in the gap, making the spectrum appear more
random and two-dimensional.

\section{Weyl formula}
\label{sec:Weyl}

As discussed in Sec.~\ref{sec:mm2txt}, the overall maximum growth rate for
the $l = 0$ and $1$ modes (and, we assume, for all $l$) occurs at $m =
\infty$.  Thus the threshold value when a given mode $l$ first starts
contributing to the spectrum is at $\lambda = -\gamma_l^2$, where
$\gamma_l$ is the maximum over $\mu$ of $\gamma(\mu,l)$.  We denote
the corresponding value of $\mu$ by $\mu_l$.

For fixed $l$ and large $\mmax$ the number of eigenvalues $N_l(\mu)$ in
an interval of $n/m$ between $\mu_l$ and $\mu$ is asymptotically equal
to the area in the $m,n$ plane [see \fig{fig:mnlattice}(a)] of the
triangle bounded by the lines $n = \mu m$, $n = \mu_l m$ and $m =
\mmax$.  That is, $N_l(\mu) \sim \half |\mu - \mu_l|\mmax^2$.

Since contours of constant $\lambda$ (or $\gamma$) asymptote to lines
of constant $\mu$ as $m \rightarrow \infty$ we can estimate the number
of eigenvalues between two values of $\lambda$ (or $\gamma$) by
inverting the function $\lambda_{\mu,l}$ for $\mu$ and substituting
this into the above expression for $N_l(\mu) $.  The inverse is
double-valued: $\mu = \mu_l^{+}(\lambda) > \mu_l$ and
$\mu_l^{-}(\lambda) < \mu_l$.  Then the number of eigenvalues between
the ground state $\lambda_{\mu_0,l}$ and $\lambda$ is approximately
\begin{equation}
		\overline{N}_l^{\pm}(\lambda)
		\equiv \half |\mu^{\pm}(\lambda) - \mu_0|\mmax^2 \;.
		\label{eq:Weyl_l}
\end{equation}

The asymptotic dependence of the total spectrum $S \equiv 
S_0^{-}\cup S_0^{+}\cup S_1^{-}\cup S_1^{+}\cup \ldots$ is thus
\begin{equation}
		\overline{N}(\lambda)  \equiv
		\sum_{l=0}^{\infty}\sum_{\pm} \overline{N}_l^{\pm}(\lambda) \;.
		\label{eq:Weyl_tot}
\end{equation}
This is the analog of the Weyl formula \cite[p. 258]{gutzwiller90} for the integral of
the smoothed spectral density (``density of states'').

Approximating $\lambda_{\mu,l} = \lambda_l + \half
(\partial^2\lambda_{\mu_l,l}/\partial \mu_l^2)(\mu-\mu_l)^2$ we get
$\mu_l^{\pm}(\lambda) = \mu_l \pm
\sqrt{2}(\lambda-\lambda_l)^{1/2}/(\partial^2\lambda_{\mu_l,l}/\partial
\mu_l^2)^{1/2}$.  Thus there is a square-root singularity at each
mode threshold.

\begin{figure}[tbp]
		\begin{tabular}{cc}
				\includegraphics[scale=0.45]{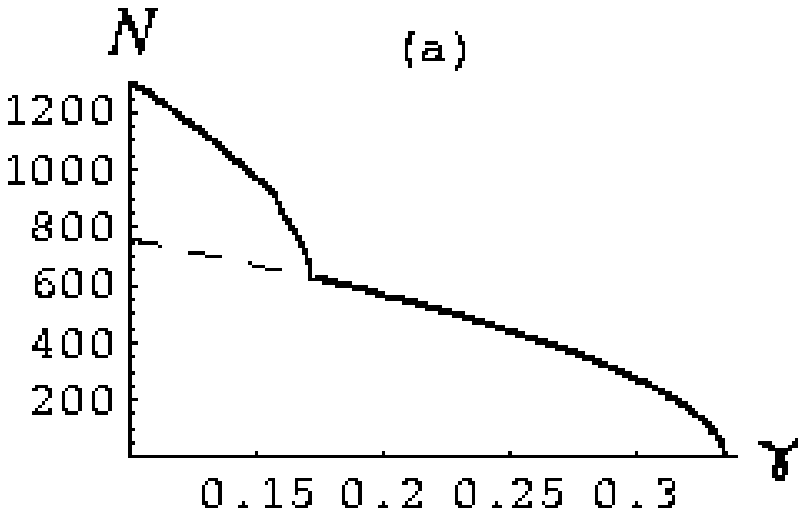} &
				\includegraphics[scale=0.45]{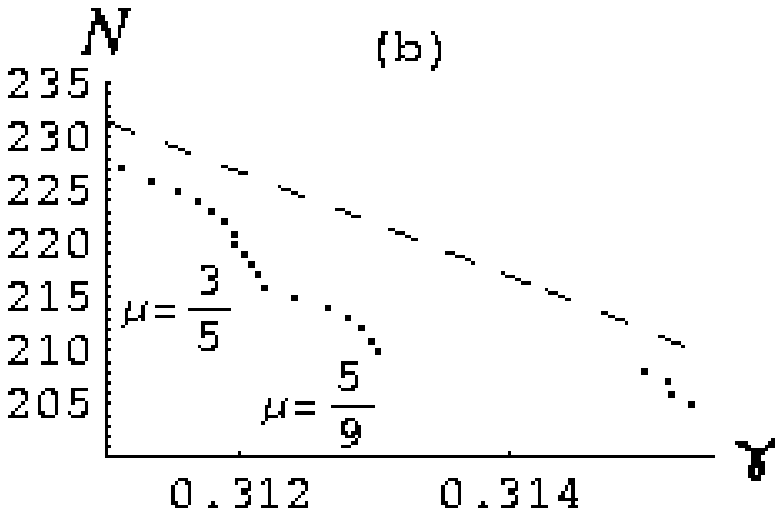} 
		\end{tabular}
		\caption{(a) Eigenvalue sequence number $N(\gamma)$ for the combined
		spectral set $S_0\cup S_1$ in the case $\mmax = 100$.  The Weyl formula
		for $S_0$,
		$\overline{N}_0^{-}(\gamma)+\overline{N}_0^{+}(\gamma)$, is shown
		dashed.  (b) Closeup of region containing eigenvalues associated 
		with $\mu = 3/5$ and $\mu = 5/9$.}
		\label{fig:NPlots}
\end{figure}

A comparison between the Weyl formula for $S_0$ and the set of points
$\{(\gamma_{N},N)\}$, where $N$ is the sequence number obtained by
sorting the set of $l=0$ and $l=1$ growth-rate eigenvalues from
largest to smallest, is shown in \fig{fig:NPlots}(a), showing
excellent agreement above the threshold for $S_1$.  The plotted points
may also be regarded as the locations of the steps in the ``staircase
plot'' of the piecewise-constant integrated density of states function
$N(\gamma)$, but the scale in this plot is too coarse to resolve the
staircase structure.

A finer-scale plot is shown in \fig{fig:NPlots}(b), in which
significant deviations from the Weyl curve are seen in the
microstructure.  The range shown in \fig{fig:NPlots}(b) is unusual in
that it contains \emph{two} well-defined accumulation sequences in
close proximity.  These are associated with low-order values of $\mu$
occurring on either side of the growth-rate maximum near $\mu = 11/19
\approx 0.579$---the sequence associated with $\mu = 5/9 \approx 0.556
$ is in $S^{-}_0$ and the one associated with $\mu = 3/5 = 0.6$ is in
$S^{+}_0$.  There are very few eigenvalues associated with high-order
rational values of $\mu$ in the range shown and the two low-order
sequences present are practically unmixed, either with each other or
with eigenvalues associated with unrelated higher-order rational
values of $\mu$.  [In fact there is only one such high-order mode in
the region of the accumulation sequences, $\mu = 51/92$, the closest
approximant to $5/9$ in the set corresponding to $S_0(1/2|5/9)$, which
causes the slight jump seen in the $\mu=3/5$ sequence.]  Also, the
wide gap containing no eigenvalues is because the intersection of the
gaps associated with the two low-order rationals is non-empty.

The spectrum near the marginal stability point, $\gamma = 0$, will
involve the superposition of many branches of radial eigenvalue $l$.
To estimate the asymptotic behavior of $N(\gamma)$ as $\gamma
\rightarrow 0$ we use the approximate dispersion relation
\eq{eq:CR4.5}.  Taking $l$ to be large we see from \eq{eq:CR4.5} that
the Suydam growth rates $\gamma_l(\mu)$ are sharply peaked about the
location of the maximum, $\mu_0$, of $G(r_{\mu})$, where $\sigma(\mu)
\equiv (G-1/4)^{1/2}$ is also a maximum.  Thus we can 
expand $\sigma(\mu)$ about $\mu_0$
\begin{equation}
         \sigma(\mu) = \sigma_{\rm max}\left[1 - 
         \left(\frac{\mu-\mu_0}{\Delta\mu}\right)^2
				 \right] + O\left((\mu-\mu_0)^3\right) \;,
\label{eq:sigmaexp}
\end{equation}
where $(\Delta\mu)^2 \equiv -2\sigma_{\rm max}/\sigma''(\mu_0)$.  To
leading order all other parameters are evaluated at the maximum point
$\mu = \mu_0$.  The quadratic correction to $\sigma$ need only be
retained in the term involving the expansion parameter $l$, so, to
leading order,
\begin{equation}
		\gamma_l \approx \gamma_0\exp 
		\left[ -\frac{\pi l}{2\sigma(\mu)} \right] \;,
		\label{eq:gammal}
\end{equation}
where $\gamma_0 \equiv 4\iotadot(r_{\mu_0})(\sigma_{\rm 
max}/e)\exp(-\pi/4\sigma_{\rm max}-1/4\sigma_{\rm max}^2)$.

Solving for $\mu$ we find
\begin{equation}
		\mu^{\pm}_l(\gamma)
		= \mu_0 \pm
		\Delta \mu 
		\left[1 - \frac{l}{l_{\rm max}(\gamma)} \right]^{1/2} \;,
		\label{eq:muasymp}
\end{equation}
where $l_{\rm max}(\gamma)\equiv (2/\pi)\sigma_{\rm max}\ln
(\gamma_0/\gamma)$.  Substituting \eq{eq:muasymp} in \eq{eq:Weyl_tot}
and approximating the sum over $l$ by an integral we find the leading
order asymptotic behavior of the number of eigenvalues to be
\begin{equation}
		\overline{N}(\gamma)  \sim
		\frac{4\Delta\mu}{3\pi}\sigma_{\rm max}\mmax^2 
		\ln\frac{\gamma_0}{\gamma} \;,
		\label{eq:Weyl_asymp}
\end{equation}
which diverges logarithmically as $\gamma \rightarrow 0$.

\begin{figure}[tbp]
		\begin{tabular}{cc}
				\includegraphics[scale=0.45]{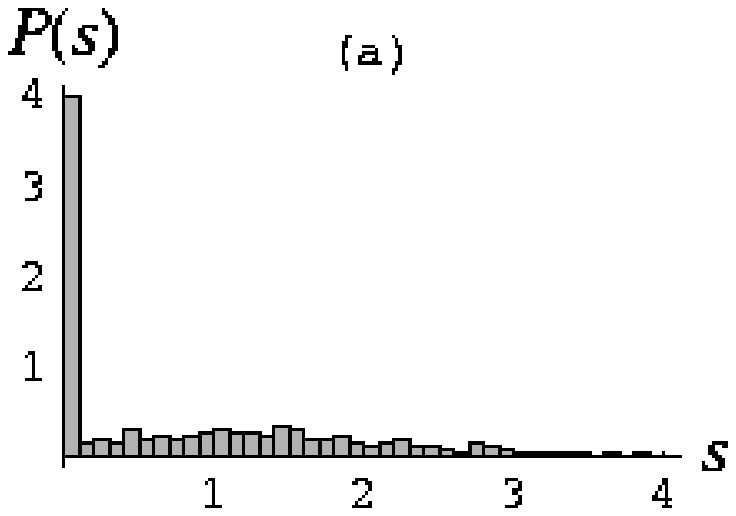} &
				\includegraphics[scale=0.45]{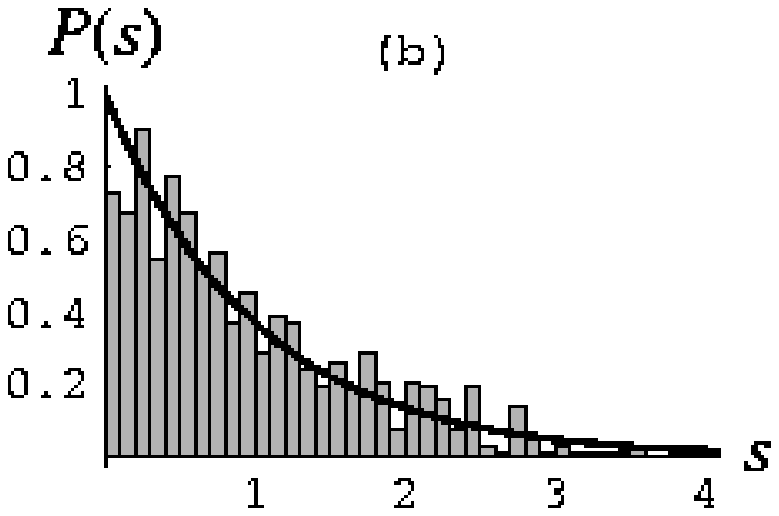} 
		\end{tabular}
		\caption{(a) Nearest-neighbor eigenvalue spacing distribution for
		the approximate spectral set $S_0^{\rm Suydam}$ using $\mu_{\rm
		min} = 0.5044$, $\mu_{\rm max} = 0.6288$, $\mmax = 100$ (625
		eigenvalues).  (b) The same, for the corresponding set of accurate eigenvalues
		$S_0$.}
		\label{fig:P100}
\end{figure}

\section{Nearest-neighbor statistics}
\label{sec:sDist}
Preparatory to the statistical analysis of eigenvalue spacing it is 
standard practice to rescale, or \emph{unfold}, the eigenvalues so as 
to make their average separation unity, thus making possible the 
comparison of different systems on the same footing.

We can unfold the spectra by using the Weyl formulae above, e.g.
for $\lambda_i \in S_0^{\pm}$ we can define rescaled eigenvalues $E_i^{\pm}$ by
\begin{equation}
		E_i^{\pm} \equiv \overline{N}_0^{\pm}(\lambda_i) \;.
		\label{eq:Unfold0pm}
\end{equation}
For the set $S_0 = S_0^{+}\cup S_0^{-}$ we can unfold with the combined
Weyl function, $\sum_{\pm}\overline{N}_0^{\pm}$.  However, for practical
purposes we have in this section used empirical least-square fits of
$N(\gamma)$ to a linear superposition of the basis functions
$(\gamma_{\rm max}-\gamma)^{1/2}$, $(\gamma_{\rm max}-\gamma)$,
$(\gamma_{\rm max}-\gamma)^{3/2}$, which captures the square-root 
singularity but avoids having to invert $\gamma_{\mu,l}$.

When $\mmax$ is large, the great majority of eigenvalues
$\lambda_{l,m,n}$ are very close to the corresponding $m=\infty$
eigenvalue with the same $\mu \equiv n/m$, $\lambda_{\mu,l}$.  Thus
one might suppose that the statistics of the spectrum are
asymptotically the same as those of an ensemble $S_0^{\rm Suydam}
\equiv \{\lambda_{n/m,0}|1 \leq m \leq \mmax, m\mu_{\rm min} < n <
m\mu_{\rm max}\}$.

In \fig{fig:P100}(a) we show the distribution of nearest-neighbor
unfolded eigenvalue spacings for $S_0^{\rm Suydam}$, and in
\fig{fig:P100}(b) that for the set $S_0$ with the correct finite-$m$
eigenvalues.  It is seen that the two distributions are radically
different---even though low-order rational values of $\mu$ are rare and
the distribution is coarse-grained, the high-$m$ approximation induces
sufficient extra degeneracy that the Suydam spectrum is dominated by a
large, but spurious, delta-function-like spike at $s=0$.  (The range of
$\mu$ used in \fig{fig:P100} corresponds to the range of $l=0$ growth
rates above the maximum $l=1$ rate, in which $S_0$ is the only
contributor to the spectrum.)

The reason why finite-$m$ effects are so important, despite the
smallness of the $O(1/m^2)$ corrections found in Sec.~\ref{sec:mm2txt},
is seen from the Weyl formula, \eq{eq:Weyl_l}, which shows that the
\emph{average} eigenvalue spacing in a set containing all values of
$n/m$ within the range of interest scales as $\mmax^{-2}$, which is the
same order as the \emph{smallest} $O(1/m^2)$ correction within a set
containing only $n/m = \const$.  Thus in the set of accurate
eigenvalues $S_0$ there is a strong intermingling of eigenvalues with
different $n/m$ that does not occur in the approximate set $S_0^{\rm
Suydam}$.

This explains why the nearest-neighbor eigenvalue spacing distribution
in \fig{fig:P100}(b) is much closer to the Poisson distribution
$\exp(-s)$ obtained for a random distribution of numbers on the real
line, and also predicted for generic separable systems
\cite{berry-tabor77}, than that in \fig{fig:P100}(a).  Nevertheless the
set of $625$ eigenvalues used in \fig{fig:P100}(b) is too small to say
convincingly that the distribution is or is not Poissonian, so we need
to analyze larger data sets to determine how close to generic the
ideal-MHD spectrum is.

\begin{figure}[tbp]
		\begin{tabular}{cc}
				\includegraphics[scale=0.45]{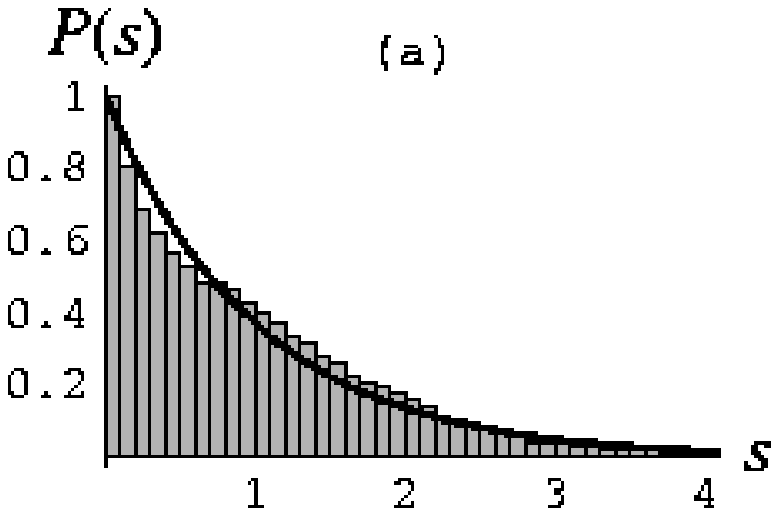} &
				\includegraphics[scale=0.45]{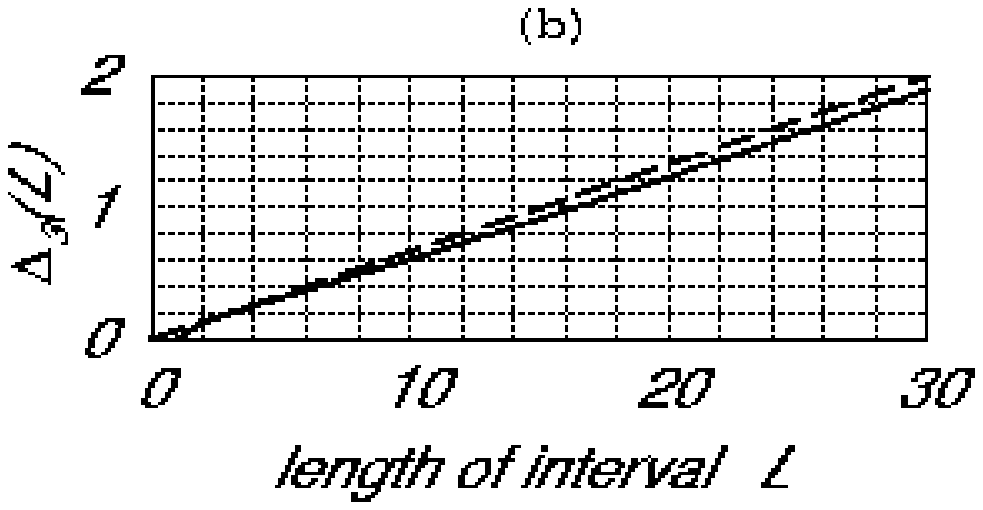} 
		\end{tabular}
		\caption{(a) Nearest-neighbor eigenvalue spacing distribution for
		the spectral set $S_0$ using $\mu_{\rm
		min} = 0.5044$, $\mu_{\rm max} = 0.6288$, $\mmax = 1000$ (62,254
		eigenvalues).  (b) The Dyson-Mehta spectral rigidity for this set 
		(solid line) compared with that for the Poisson process (dashed line).}
		\label{fig:bigset}
\end{figure}

A cutoff at $\mmax=1000$ gives a set $S_0$ containing about $62,254$
eigenvalues in the range between the maximum $l=0$ growth rate and the
maximum $l=1$ growth rate.  [Note the approximately $\mmax^2$ scaling
in the size of $S_0$, as predicted by the Weyl formula,
\eq{eq:Weyl_l}.]  In \fig{fig:bigset}(a) we show the nearest-neighbor
distribution for this set.  Close examination of the region near the
origin reveals no trace of the spike seen in \fig{fig:P100}(a), not
even the tiny spike found by Casati \emph{et al.}
\cite{casati-chirikov-guarneri85} for the spectrum of waves in an
incommensurate rectangular box.  However, it is clear that the
statistics are not exactly Poissonian.

In \fig{fig:bigset}(b) we show the Dyson-Mehta rigidity parameter
$\Delta_3(L)$ \cite[pp.  321--323]{mehta91}, defined as the
least-squares deviation of the unfolded eigenvalue staircase $N(E)$
from the best-fitting straight line in an interval of length $L$.
Again, the behavior is similar to that for the completely random
spectrum (Poisson process) in that $\Delta_3$ increases linearly with
$L$, but the slope is slightly less than the $1/15$ expected for the
Poisson processs.

\begin{figure}[tbp]
		\begin{tabular}{cc}
				\includegraphics[scale=0.45]{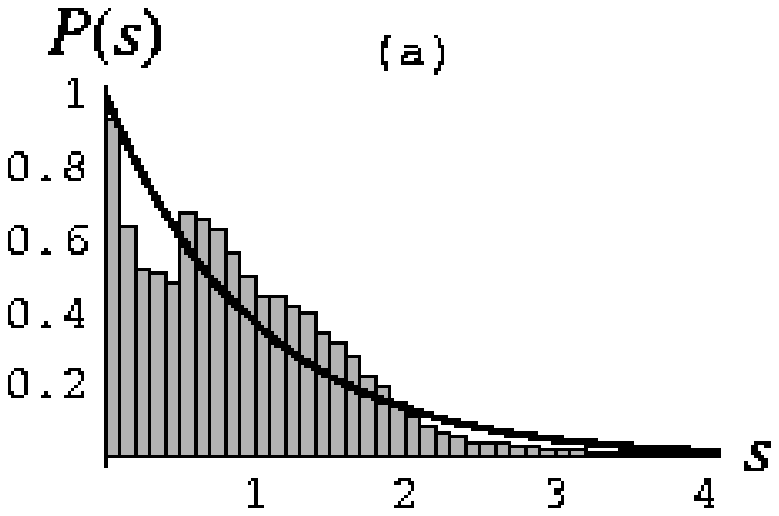} &
				\includegraphics[scale=0.45]{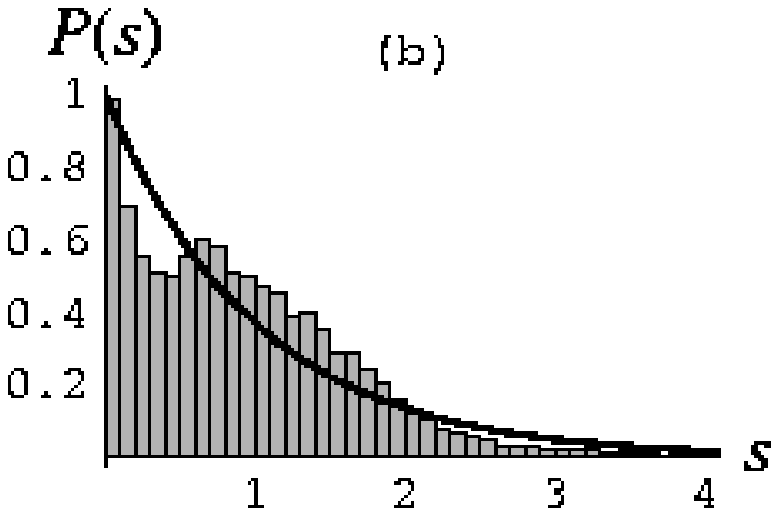} 
		\end{tabular}
		\caption{(a) Nearest-neighbor eigenvalue spacing distribution for
		the spectral set $S_0^{-}$ for $\mmax = 1000$ (37,932
		eigenvalues).  (b) Nearest-neighbor eigenvalue spacing distribution for
		the spectral set $S_0^{+}$ (24,412 eigenvalues).}
		\label{fig:bigsetpm}
\end{figure}

In order to understand the departure from Poisson statistics better,
we show in \fig{fig:bigsetpm} the spacing distribution for the
corresponding sets $S_0^{-}$ and $S_0^{+}$.  The departure from
Poisson statistics is now quite striking.  This is presumably because
of the gaps about low-order rational values of $\mu$ mentioned in
Sec.~\ref{sec:S0}, which leave the 1-dimensional accumulation
sequences $S_0(\mu|\mu)$ unmixed with other parts of the spectrum, so
the spacing distribution combines aspects of that for a 1-dimensional
system (peaked at 1) and that for a generic separable 2-dimensional
system (peaked at 0).

\begin{figure}[tbp]
		\begin{tabular}{cc}
				\includegraphics[scale=0.45]{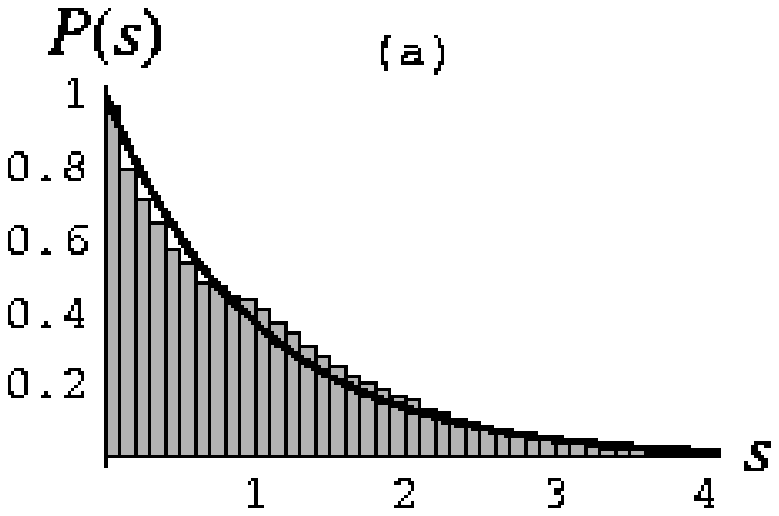} &
				\includegraphics[scale=0.45]{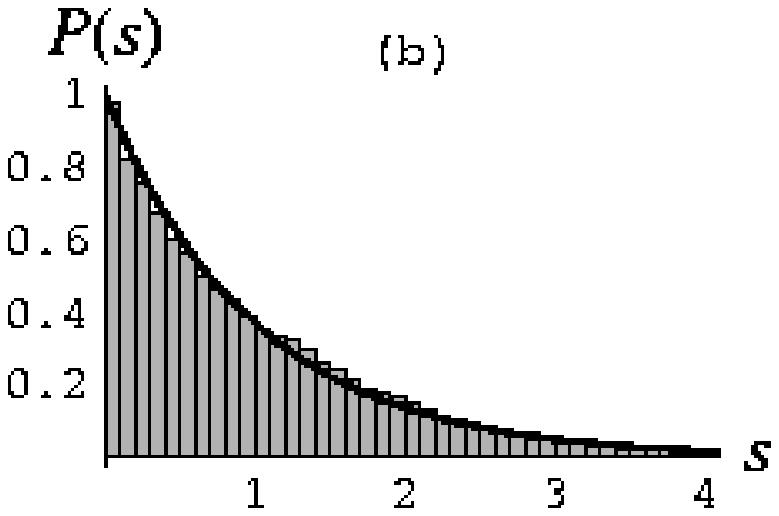} 
		\end{tabular}
		\caption{(a) Nearest-neighbor eigenvalue spacing distribution for
		the first $72,500$ eigenvalues of the $l=1$ spectral set $S_1$,
		$\mmax = 1000$.  (b) Nearest-neighbor eigenvalue spacing
		distribution for the mixed spectral set $S_0 \cup S_1$ over the 
		same range of eigenvalues as in (a) (total of 95,000 eigenvalues).}
		\label{fig:bigset01}
\end{figure}

In \fig{fig:bigset01}(a) we show the spacing distribution for the $l =
1$ spectrum, which is seen to be very much like the $l = 0$ spectrum
of \fig{fig:bigset}(a) in its departure from the Poisson
distribution.  However, we might expect that mixing the $l = 0$ with
the $l = 1$ spectrum will make the levels appear more ``random'' and
\fig{fig:bigset01}(b) confirms that the level spacing distribution
does indeed become more like the exponential expected for a Poisson
process.

\section{Conclusion}

We have demonstrated that the statistical nature of the ideal-MHD
interchange spectrum deviates significantly from the random Poisson
process of generic separable systems due to the number-theoretic
structure of the eigenvalue distribution.  The similarity between the
two level-spacing distributions in \fig{fig:bigsetpm}, which
correspond to two different parts of the rotational transform profile,
suggest the possibility that there may nevertheless be some
universality in the statistics. If so, we have found a new 
universality class.

The crude regularization used in this paper, simply restricting the
poloidal mode numbers to $m \leq\mmax$, is not very physical but
corresponds closely to what is done in the large three-dimensional
eigenvalue codes CAS3D \cite{schwab93} and TERPSICHORE
\cite{anderson_etal90}.  Thus, apart from fundamental mathematical
interest, the primary motivation of this paper has been the numerical
analysis of the three-dimensional ideal-MHD spectrum as produced by
these codes.  Preliminary results \cite{dewar-nuehrenberg-tatsuno04}
on an interchange-unstable stellarator test case show spectra
with eigenvalue separation statistics similar to those of strongly
quantum chaotic systems.  However, the results of the present paper
indicate that some caution should be taken in interpreting ideal-MHD
spectra in terms of conventional quantum chaos theory because of the
radically different nature of the dispersion relation.

In subsequent work it will be important to examine the effect of
finite Larmor radius on the spectrum.  However, this typically makes the
problem non-Hermitian and less easy to compare with standard quantum
chaos theory.

\begin{acknowledgments}
One of us (RLD) acknowledges the support of the Australian Research
Council and useful discussions with H. Friedrich, R. Mennicken, H.
Schomerus, G. Spies, J. Wiersig and N. Witte, on spectral and quantum
chaos issues forming the background of this paper.
 
\end{acknowledgments}

\appendix
\section{$1/m^2$ Corrections}
\label{sec:mm2app}

The coefficients of the expansion \eq{eq:Expansion} are found by 
Taylor expansion of the geometric and equilibrium quantities in 
Eqs.~(\ref{eq:Ldef})
\begin{eqnarray}
	L^{(0)} & =  &  -\frac{d}{dx}\iotadot^2 x^2\frac{d}{dx} 
     	+\iotadot^2 x^2 - \Ds \;, \nonumber \\
	L^{(1)} & =  & x\frac{d}{dx}\iotadot^2 x^2\frac{d}{dx} 
	                     -\frac{d}{dx}\iotadot\,\iotaddot\, x^3\frac{d}{dx}
											 \nonumber \\  & & \mbox{}
					+\iotadot\,(\iotaddot-3\,\iotadot)x^3 + 
					(2\Ds - \Dsdot- \iotadot\,\iotaddot) x \;, \nonumber \\
	L^{(2)} & =  & -x^2\frac{d}{dx}\iotadot^2 x^2\frac{d}{dx} 
	                     +x\frac{d}{dx}\iotadot\,\iotaddot\, x^3\frac{d}{dx} 
											 \nonumber \\  & & \mbox{}
					   + \frac{d}{dx} x^4\left(
						            \frac{\iotadot^2}{12}  - \frac{\iotaddot^2}{4}
								+ \frac{\iotadot\,\iotaddot}{2} - \frac{\iotadot\,\iotadddot}{3}
						 \right)\frac{d}{dx} 
											\nonumber \\  & & \mbox{}
					   +\frac{x^2}{2}\left(
								5\,\iotadot\,\iotaddot - 2\,\iotadot\,\iotadddot - \iotaddot^2
							   - 6\Ds +5\Dsdot - \Dsddot
						                     \right) 
									 		\nonumber \\  & & \mbox{}
					   +x^4\left(
						            \frac{71\,\iotadot^2}{12}  + \frac{\iotaddot^2}{4}
								- \frac{7\,\iotadot\,\iotaddot}{2} + 
								\frac{\iotadot\,\iotadddot}{3}
					      	 \right)
   \label{eq:Lexp} 
\end{eqnarray}
and (\ref{eq:Mdef}) 
\begin{eqnarray}
	M^{(0)} & = & -\frac{d^2}{dx^2} + 1 \;, \nonumber \\
	M^{(1)} & = & x\frac{d^2}{dx^2}  -  \frac{d}{dx}x\frac{d}{dx}  - 2x \;, \nonumber \\
	M^{(2)} & = & -x^2\frac{d^2}{dx^2}  +  x\frac{d}{dx}x\frac{d}{dx} + 
	3x^2 \;.
	 \label{eq:Mexp}
\end{eqnarray}
where, as in the main text, dots denote nondimensional derivatives,
$\iotadot \equiv \rmu d\iota/dr$ etc., and all equilibrium quantities
are evaluated at $r = \rmu$.

% \bibliography{Ballooning}% Produces the bibliography via BibTeX.

\end{document}